\documentclass[a4paper]{jpconf}
\usepackage{graphicx}

\begin{document}

\title{Nonequilibrium work performed in quantum annealing}
\author{Masayuki Ohzeki, and Hidestoshi Nishimori}

\address{Department of Systems Science, Graduate School of Informatics, Kyoto University, Yoshida Honmachi, Sakyo-ku, Kyoto, 606-8501, Japan\\
Department of Physics, Tokyo Institute of Technology, Oh-okayama, Meguro-ku, Tokyo, 152-8551, Japan}

\ead{mohzeki@i.kyoto-u.ac.jp}

\begin{abstract}
Quantum annealing is a generic solver of classical optimization
problems that makes full use of quantum fluctuations. We consider work statistics
given by a repetition of quantum annealing processes by employing the Jarzynski equality
proposed in nonequilibrium statistical physics. 
In particular, we analyze a nonequilibrium average of the exponentiated work performed by a transverse field.
A special symmetry, gauge symmetry, leads to a non-trivial relationship between
quantum annealing toward different targets in the theory of spin glasses.
We believe that our results will be a step toward an alternative
realization of efficient quantum computation as well as our better understanding of
nonequilibrium behavior of systems under quantum control.
\end{abstract}

\section{Introduction}

We should often solve difficult problems in a reasonable time, which is
closely related to the efficiency in cost and time in industry and distribution systems.
A well-known instance is to find the
minimum path for a person to visit all the cities, called the traveling
salesman problem, which is a typical optimization problem \cite{OP,OP2}.
The goal of quantum computation is to solve problems hard to solve quickly
on classical computer in a moderate time
by use of superpositions, tunneling effects, and entanglement of
quantum nature. Quantum annealing is a method to realize quantum computation \cite{QA1,QA2,QA3,QA4}.
Conventional quantum computation uses successive, discrete
unitary operations to enhance the probability of the desired state representing
the answer of a hard optimization problem. In contrast, quantum annealing
uses continuous dynamics represented by a time-dependent Schr\"odinger equation.
Dynamical behavior of quantum systems during such processes is still poorly
understood, and an adiabatic control of quantum dynamics is often employed,
called the quantum adiabatic computation \cite{QAA}. Such a quantum adiabatic
computation, however, has a bottleneck for a certain type of the
optimization problems \cite{FT1,FT2}. 

A recent study of quantum annealing indicates a possible deviation
from adiabatic controls \cite{QJA}. In the present paper, we proceed along this line
and analyze the work performed by quantum fluctuations in a non-adiabatic control of system
parameter. The
analysis makes use of several remarkable properties found in statistical physics.
One is an exact relationship between nonequilibrium processes and equilibrium
states at the beginning and end, the Jarzynski equality \cite{JE1,JE2}.
Another is a special symmetry, called the gauge symmetry, found in spin
glasses \cite{HN81,HNbook}.
As detailed below, a combination of these theories enables us to
obtain exact relations between equilibrium and nonequilibrium quantities
and to evaluate several quantities related to dynamical processes in quantum systems \cite{JPSJ2010}.

After an introduction of the Jarzynski equality,
we develop our analyses by use of gauge symmetry to show the derivation of an exact
identity related with the performed work during quantum annealing.

\section{Quantum annealing and spin glasses}

We consider the following time-dependent Hamiltonian in quantum annealing 
\begin{equation}
H(t)=-f(t)J\sum_{\langle ij\rangle }\tau _{ij}\sigma _{i}^{z}\sigma
_{j}^{z}-\Gamma \left( 1-f(t)\right) \sum_{i}\sigma _{i}^{x},  \label{Ham}
\end{equation}%
where $f(t)=t/T$, and $t$ runs continuously from $0$ to $T$. The initial
Hamiltonian $H(0)$ is composed only of the transverse-field term $-\Gamma
\sum_{i}\sigma _{i}^{x}$. The final Hamiltonian $H(T)$ is chosen to describe the optimization
problem to be solved. We take a simple spin glass Hamiltonian as $H(T)$, the $\pm J$
Ising model, on an arbitrary lattice throughout this paper for convenience. The sign of the
interaction $\tau _{ij}$ follows the distribution function 
\begin{equation}
P(\tau _{ij})=p\delta (1-\tau _{ij})+(1-p)\delta (1+\tau _{ij})=\frac{%
\mathrm{e}^{\beta _{p}J\tau _{ij}}}{2\cosh \beta _{p}J},\label{distribution}
\end{equation}%
where $p$ is the concentration of the ferromagnetic interaction $J>0$, and $%
\exp (-2\beta _{p}J)=(1-p)/p$. A sufficiently slow decrease of the strength of
the transverse field changes the trivial initial state to the nontrivial
ground state of the target Hamiltonian $H(T)$. This is the idea of quantum
adiabatic computation realized in $T\to\infty$. Instead of the adiabatic
control, in the present study, let us consider the repetition of fast
quantum annealing ($T$ small) starting from an equilibrium ensemble. We may not be able to always
reach the desired state since the system does not trace the
instantaneous ground state when the adiabatic condition is not satisfied.
Therefore we need to repeat the process to hit
the correct ground state in such a non-adiabatic realization of quantum annealing.

We next point out that
a class of models of spin glasses has a special symmetry known as the gauge symmetry.
We define the gauge transformation as the following local unitary operator
for Pauli spin matrices, and the change of the sign of the interaction \cite%
{Morita}: 
\begin{equation}
\sigma_i^{\alpha} \to G \sigma_i^{\alpha} G^{-1}, G = \prod_i G_i,\quad G_i
= \left\{ 
\begin{array}{ll}
1 & (\xi_i = +1) \\ 
\exp(\mathrm{i} \pi\sigma_i^x/2) & (\xi_i = -1)%
\end{array}
\right.
\end{equation}
where $\alpha =x, y, z$
and $\tau_{ij} = \tau_{ij}\xi_i\xi_j$, where $\xi$ is a classical gauge
variable taking only $\pm 1$. After the above gauge transformation, the
time-dependent Hamiltonian (\ref{Ham}) is invariant, but the distribution
function (\ref{distribution}) is modified into $P(\tau_{ij}) = \exp(\beta_p J \tau_{ij}\xi_i
\xi_j)/2\cosh \beta_p J$. This property helps us to derive the following
results.

\section{Analysis}

\subsection{Jarzynski equality}

Let us introduce another piece of theoretical tool, the Jarzynski equality, which is
a generalization of the second law of thermodynamics. We here use the
quantum version of the Jarzynski equality for the $\pm J$ Ising model, not the
original version for the classical case \cite{JE1,JE2},  with a
specific configuration $\{\tau _{ij}\}$ as follows \cite{QJE1,QJE2},
\begin{equation}
\langle \mathrm{e}^{-\beta W}\rangle _{0\rightarrow T}=\frac{Z_{\beta
}(T;\{\tau _{ij}\})}{Z_{\beta }(0;\{\tau _{ij}\})},  \label{JE}
\end{equation}%
where $Z_{\beta }(t;\{\tau _{ij}\})$ is the partition function for the
instantaneous Hamiltonian, and $\beta $ is the inverse temperature. The
performed work during a nonequilibrium process is given by the difference
between the outputs of the measurements of the initial and final energies
as $W=E_{m}(T)-E_{n}(0)$, where $m$ and $n$ denote the indices of the eigenstates as 
$H(T)|m(T)\rangle =E_{m}(T)|m(T)\rangle $ and $H(0)|n(0)\rangle
=E_{n}(0)|n(0)\rangle $. The left-hand side of equation (\ref{JE})
expresses the
average of the exponentiated work over all the realizations of nonequilibrium
processes starting from the equilibrium ensemble, and is
evaluated as 
\begin{eqnarray}
\langle \mathrm{e}^{-\beta W}\rangle _{0\rightarrow T} &=&\sum_{m,n}\mathrm{e%
}^{-\beta (E_{m}(T)-E_{n}(0))}P(m|n)\frac{\mathrm{e}^{-\beta E_{n}(0)}}{%
Z_{\beta }(0;\{\tau _{ij}\})}  \nonumber \\
&=&\frac{1}{Z_{\beta }(0;\{\tau _{ij}\})}\sum_{m,n}\mathrm{e}^{-\beta
E_{m}(T)}P(m|n)=\frac{Z_{\beta }(T;\{\tau _{ij}\})}{Z_{\beta }(0;\{\tau
_{ij}\})},
\end{eqnarray}%
where we use the path probability for a nonequilibrium process as, $%
P(m|n)=|\langle m(T)|U_{0\rightarrow T}|n(0)\rangle |^{2}$. Here $%
U_{0\rightarrow T}$ is the time evolution operator.

\subsection{Jarzynski equality for quantum annealing : Symmetric distribution%
}

The partition function for the initial Hamiltonian $H(0)$
does not depend on the configuration $\{\tau_{ij}\}$ since it is given only
by the transverse field, $Z_{\beta}(0;\{\tau_{ij}\}) = \left( 2 \cosh
\beta\Gamma \right)^{N}.$ Here $N$ is the number of spins. Let us now
consider the configurational average of the Jarzynski equality for quantum
annealing by summation over all the possible configurations $\{\tau_{ij}\}$
as 
\begin{equation}
\left[\langle \mathrm{e}^{-\beta W} \rangle_{0 \to T}\right]_{\beta_p} = 
\frac{\left[Z_{\beta}(T;\{\tau_{ij}\})\right]_{\beta_p}}{\left( 2 \cosh
\beta\Gamma \right)^{N}} .  \label{An}
\end{equation}
The explicit expression of the quantity on the right-hand side is 
\begin{equation}
\frac{\left[Z_{\beta}(T;\{\tau_{ij}\})\right]_{\beta_p}}{\left( 2 \cosh
\beta\Gamma \right)^{N}} = \frac{1}{\left( 2 \cosh \beta\Gamma
\right)^{N}\left( 2 \cosh \beta_p J\right)^{N_B}}\sum_{\tau_{ij}}\prod_{\langle ij \rangle}\mathrm{e}%
^{\beta_p J \tau_{ij}}Z_{\beta}(T;\{\tau_{ij}\}),
\end{equation}
where $N_B$ is the number of interactions. If we set $\beta_p = 0$, which
corresponds to the case with the symmetric distribution of the interaction ($%
p=1/2$), the above quantity can be evaluated as 
\begin{equation}
\left[\langle \mathrm{e}^{-\beta W} \rangle_{0 \to T}\right]_{\beta_p=0} = 
\frac{2^N\left( 2 \cosh \beta J \right)^{N_B}}{2^{N_B}\left( 2 \cosh
\beta\Gamma \right)^{N}}.
\end{equation}

\subsection{Inverse statistics on special subspace}

Let us show that the gauge transformation reveals a non-trivial property
of the nonequilibrium average of the exponentiated work during quantum
annealing. We take the configurational average of the inverse of
the nonequilibrium-averaged exponentiated work as 
\begin{equation}
\left[ \frac{1}{\langle \mathrm{e}^{-\beta W}\rangle _{0\rightarrow T}}%
\right] _{\beta _{p}}=\frac{\left( 2\cosh \beta \Gamma \right) ^{N}}{%
\left( 2\cosh \beta _{p}J\right) ^{N_{B}}}\sum_{\tau _{ij}}\frac{\prod_{\langle ij \rangle}
\mathrm{e}^{\beta _{p}J\tau _{ij}}}{Z_{\beta }(T;\{\tau _{ij}\})}.
\end{equation}%
Here we apply the gauge transformation to the above equality. The quantity
on the left-hand side does not change, since the Hamiltonian and the
time-evolution operator are invariant. On the other hand, the
quantity on the right-hand side becomes 
\begin{equation}
\left[ \frac{1}{\langle \mathrm{e}^{-\beta W}\rangle _{0\rightarrow T}}%
\right] _{\beta _{p}}=\frac{\left( 2\cosh \beta \Gamma \right) ^{N}}{%
\left( 2\cosh \beta _{p}J\right) ^{N_{B}}}\sum_{\tau _{ij}}\frac{\prod_{\langle ij \rangle}
\mathrm{e}^{\beta _{p}J\tau _{ij}\xi _{i}\xi _{j}}}{Z_{\beta }(T;\{\tau
_{ij}\})}.
\end{equation}%
Setting $\beta =\beta _{p}$
 and taking the summation over all configurations of $\xi _{i}$ leads us to 
\begin{equation}
\left[ \frac{1}{\langle \mathrm{e}^{-\beta W}\rangle _{0\rightarrow T}}%
\right] _{\beta }=\frac{2^{N_{B}}\left( 2\cosh \beta \Gamma \right) ^{N}}{%
2^{N}\left( 2\cosh \beta J\right) ^{N_{B}}},
\end{equation}%
which implies that 
\begin{equation}
\left[ \langle \mathrm{e}^{-\beta W}\rangle _{0\rightarrow T}\right] _{\beta
_{p}=0}=\left( \left[ \frac{1}{\langle \mathrm{e}^{-\beta W}\rangle
_{0\rightarrow T}}\right] _{\beta _{p}=\beta }\right) ^{-1}.
\end{equation}%
This equality represents a non-trivial relationship in quantum annealing driven toward two
different states in spin glasses $\beta _{p}=0$ and $\beta _{p}=\beta $. It is
important to remark that we do not use any approximations in the above
calculation, and we deal with the nonequilibrium average during quantum
annealing, which does not assume to be quasi-static, unlike quantum adiabatic
computation.

\section{Summary}

We can obtain several further exact equalities for the repetition of non-adiabatic
quantum annealing and establish non-trivial relations that hold between quantum annealing
toward different spin glasses, as will be announced shortly in a separate paper. 
Exactness of our results would be useful as a benchmark
to check experimental conditions of nonequilibrium process in
quantum spin dynamics. Notice that our results are scalable, which means
independence of the system size, since the Jarzynski equality holds for any size if we take all
fluctuations into account in estimation of the expectation value on the
left-hand side of equation (\ref{JE}). Finite errors on the ratio of the partition functions may
be attributed to the rare event on nonequilibrium processes. The present study
will help us to understand and manipulate quantum dynamics efficiently.

%\section*{Acknowledgement}
%
%This work was financially supported by the 21st Century Global COE Program
%at Tokyo Institute of Technology `Nanoscience and Quantum Physics', and by
%CREST, JST.


\section*{References}

\begin{thebibliography}{99}
\bibitem{OP} Garey M R and Johnson D S 1979 \textit{Computers and Intractability: A
Guide to the Theory of NP-Completeness} (San Francisco: Freeman)

\bibitem{OP2} Hartmann A K and Weigt M 2005 \textit{Phase Transitions in
Combinatorial Optimization Problems: Basics, Algorithms and Statistical
Mechanics} (Wiley-VCH, Weinheim)

\bibitem{QA1} Kadowaki T and Nishimori H 1998 \textit{Phys. Rev. E} \textbf{%
58} 5355

\bibitem{QA2} Das A and Chakrabarti B K 2008 \textit{Rev. Mod. Phys.} \textbf{80} 1061

\bibitem{QA3} Morita S and Nishimori H 2008 \textit{J. Math. Phys.} \textbf{%
49} 125210

\bibitem{QA4} Ohzeki M and Nishimori H 2011 to appear in J. Comp. Theor.
Nanoscience

\bibitem{QAA} Farhi E, Goldstone J, Gutmann S and Sipser M, arXiv:0001106

\bibitem{FT1} J\"org T, Krzakala F, Kurchan J and Maggs A C 2008 \textit{Phys.
Rev. Lett.} \textbf{101} 147204

\bibitem{FT2} Young A P, Knysh S and Smelyanskiy V N 2010 \textit{Phys. Rev.
Lett.} \textbf{104} 020502

\bibitem{QJA} Ohzeki M 2010 \textit{Phys. Rev. Lett.} \textbf{105} 050401

\bibitem{JE1} Jarzynski C 1997 \textit{Phys. Rev. Lett.} \textbf{78} 2690

\bibitem{JE2} Jarzynski C 1997 \textit{Phys. Rev. E} \textbf{56} 5018

\bibitem{HN81} Nishimori H 1981 \textit{Prog. Theor. Phys.} \textbf{66} 1169

\bibitem{HNbook} Nishimori H \emph{Statistical Physics of Spin Glasses and
Information Processing: An Introduction} (Oxford Univ. Press, Oxford, 2001)

\bibitem{JPSJ2010} Ohzeki M and Nishimori H 2010 \textit{J. Phys. Soc, Jpn.} \textbf{79} 084003

\bibitem{Morita} Morita S, Nishimori H and Ozeki Y 2006 \textit{J. Phys.
Soc. Jpn.} {\textbf{7}5} 014001

\bibitem{QJE1} Tasaki H arXiv:0009244

\bibitem{QJE2} Campisi M,  Talkner P and H\"anggi P 2009 \textit{Phys. Rev. Lett.} \textbf{102} 210401
\end{thebibliography}
\end{document}